\documentclass[sigconf,natbib=true]{acmart}
\usepackage{booktabs}
\usepackage{ragged2e}
\usepackage{nicematrix}
\usepackage{tabularx}
\usepackage{adjustbox}

\usepackage{graphicx}

\usepackage{enumitem}

\usepackage{colortbl}
\usepackage{multirow}
\usepackage{graphicx}
\usepackage[normalem]{ulem}
\useunder{\uline}{\ul}{}

\usepackage{algorithm}
\usepackage{algpseudocode}

\definecolor{gblue}{RGB}{66,133,244}

\usepackage{arydshln}
\makeatletter
\def\adl@drawiv#1#2#3{%
        \hskip.5\tabcolsep
        \xleaders#3{#2.5\@tempdimb #1{1}#2.5\@tempdimb}%
                #2\z@ plus1fil minus1fil\relax
        \hskip.5\tabcolsep}
\newcommand{\cdashlinelr}[1]{%
  \noalign{\vskip\aboverulesep
           \global\let\@dashdrawstore\adl@draw
           \global\let\adl@draw\adl@drawiv}
  \cdashline{#1}
  \noalign{\global\let\adl@draw\@dashdrawstore
           \vskip\belowrulesep}}
\makeatother

\usepackage[normalem]{ulem} %

\usepackage{colortbl} %
\definecolor{mygray}{rgb}{0.9, 0.9, 0.9}
\definecolor{myred}{rgb}{0.68627451, 0.14117647, 0.09803922}

\newcommand{\mytab}[1]{{Table~\ref*{tab:#1}}}
\newcommand{\myfig}[1]{{Fig.~\ref*{fig:#1}}}

\AtBeginDocument{%
  }

\copyrightyear{2025} 
\acmYear{2025} 
\setcopyright{acmlicensed}
\acmConference[Conference acronym 'XX]{Make sure to enter the correct
  conference title from your rights confirmation email}{June 03--05,
  2025}{Woodstock, NY}
\acmDOI{XXXXXXX.XXXXXXX}
\acmISBN{978-1-4503-XXXX-X/2025/11}


\begin{document}

\title{MindRec: A Diffusion-driven Coarse-to-Fine Paradigm for Generative Recommendation} %

\author{Mengyao Gao}
\email{mengyao0301@mail.ustc.edu.cn}
\orcid{0009-0004-5517-3563}
\affiliation{%
  \institution{University of Science and Technology of China}
  \city{Hefei}
  \country{China}
}

\author{Chongming Gao}
\authornote{Corresponding Author.}
\email{chongminggao@ustc.edu.cn}
\affiliation{%
  \institution{University of Science and Technology of China}
  \city{Hefei}
  \country{China}
}

\author{Haoyan Liu}
\email{liuhaoyan@ustc.edu.cn}
\affiliation{%
  \institution{University of Science and Technology of China}
  \city{Hefei}
  \country{China}
}

\author{Qingpeng Cai}
\email{cqpcurry@gmail.com}
\affiliation{%
  \institution{Independent}
  \city{Beijing}
  \country{China}
  \country{}
}

\author{Peng Jiang}
\email{jp2006@139.com}
\affiliation{%
  \institution{Independent}
  \city{Beijing}
  \country{China}
  \country{}
}

\author{Jiajia Chen}
\email{jia2chan@mail.ustc.edu.cn}
\affiliation{%
  \institution{Anhui University}
  \city{Hefei}
  \country{China}
}

\author{Shuai Yuan}
\email{syuanaf@connect.ust.hk}
\orcid{0000-0001-6730-5755}
\affiliation{%
  \institution{Hong Kong University of Science and Technology}
  \city{Hong Kong}
  \country{China}
  \country{}
}

\author{Xiangnan He}
\authornotemark[1]
\email{xiangnanhe@gmail.com}
\orcid{0000-0001-8472-7992}
\affiliation{%
  \institution{MoE Key Lab of BIPC, University of Science and Technology of  China}
  \city{Hefei}
  \country{China}
  \country{}
}

\renewcommand{\shortauthors}{Mengyao Gao et al.}

\begin{abstract}
Recent advancements in large language model-based recommendation systems often represent items as text or semantic IDs and generate recommendations in an auto-regressive manner. However, due to the left-to-right greedy decoding strategy and the unidirectional logical flow, such methods often fail to produce globally optimal recommendations. In contrast, human reasoning does not follow a rigid left-to-right sequence. Instead, it often begins with keywords or intuitive insights, which are then refined and expanded.
Inspired by this fact, we propose MindRec, a diffusion-driven coarse-to-fine generative paradigm that emulates human thought processes. Built upon a diffusion language model, MindRec departs from auto-regressive generation by leveraging a masked diffusion process to reconstruct items in a flexible, non-sequential manner. Particularly, our method first generates key tokens that reflect user preferences, and then expands them into the complete item, enabling adaptive and human-like generation. To further emulate the structured nature of human decision-making, we organize items into a hierarchical category tree. This structure guides the model to first produce the coarse-grained category and then progressively refine its selection through finer-grained subcategories before generating the specific item. To mitigate the local optimum problem inherent in greedy decoding, we design a novel beam search algorithm, Diffusion Beam Search, tailored for our mind-inspired generation paradigm.
Experimental results demonstrate that MindRec yields a 9.5\% average improvement in top-1 accuracy over state-of-the-art methods, highlighting its potential to enhance recommendation performance. The implementation is available via https://github.com/Mr-Peach0301/MindRec.

\end{abstract}


\maketitle

\section{Introduction}

\begin{figure}[!t]
  \centering
  \includegraphics[width=\columnwidth]{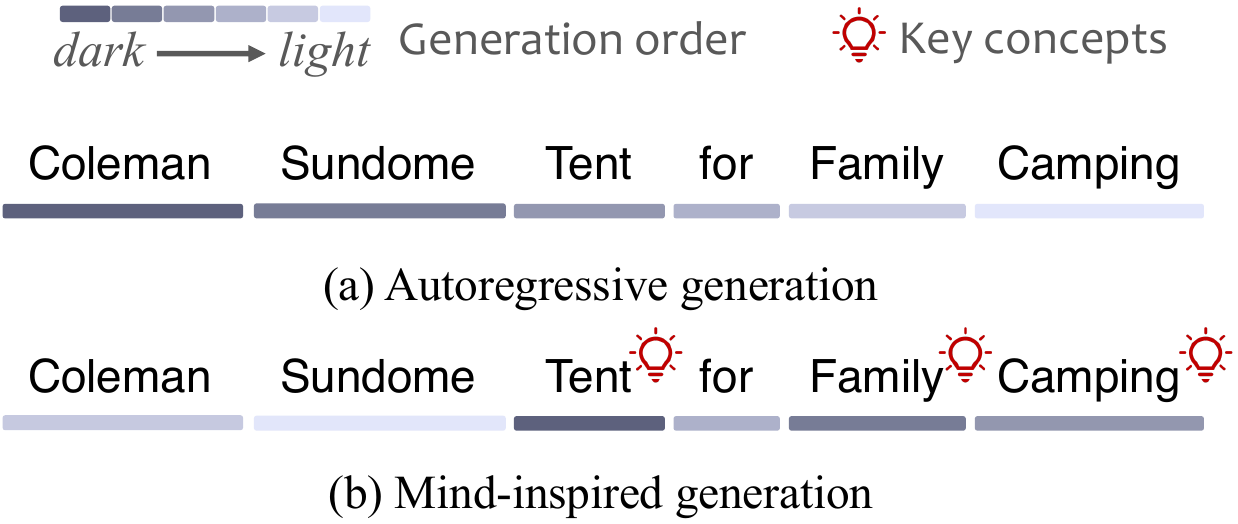} %
  \caption{Illustration of two decoding paradigms of generative
recommendation.} %
  \label{fig:paradigm} %
\end{figure}

Large Language Models (LLMs) have recently gained traction in recommendation systems, owing to their extensive world knowledge and powerful representation capabilities \cite{surveyLLM4rec,Zhao_2024,li2024generative}. A typical formulation is next-item prediction: given a sequence of user interactions, the goal is to predict the next item with which the user is likely to engage \cite{jiang2024item,bao2024decoding}. To this end, prior work has explored representing items either as natural language text \cite{gao2024sprec,bao2023tallrec,zhang2023instructrec} or as semantic identifiers that maintain fixed length while better capturing collaborative or semantic signals \cite{wang2024learnable,zheng2024adapting,rajput2023tiger}. 
Training typically follows the standard left-to-right objective: the model maximizes the likelihood of the next target token given its preceding tokens \cite{gao2025flower}. At inference time, tokens are generated auto-regressively, with each step selecting the most probable token based on previously generated context.

While effective in many domains, the auto-regressive nature of LLMs poses a fundamental limitation for recommendations. In particular, unidirectional and greedy decoding strategies often lead to locally optimal sequences and fail to reflect users' intent \cite{lin2025order,10.5555/3037062.3037074}. The core issue is that generation proceeds strictly token by token: if the first token of a relevant target item is assigned a low probability, the item is prematurely eliminated from the candidate space, regardless of its overall relevance. Once excluded at this early stage, no subsequent evidence can recover it, making the system highly sensitive to local probability fluctuations. This structural weakness represents one of the most critical challenges of applying auto-regressive LLMs to recommendation tasks.

In contrast to the auto-regressive paradigm, human reasoning does not strictly proceed in a left-to-right sequence. Rather, it often begins by retrieving salient keywords or core insights, which are then iteratively expanded into a coherent response \cite{duncker1945problem, Guilford1950Creativity}. For instance, as shown in \myfig{paradigm}, when recommending the next item, a human might first extract a concept such as ``Tent'' from recent engagements with outdoor gear. This cue is then placed within the broader user profile, leading to the inference that the user is likely someone with a family seeking ``Family Camping'' products. The reasoning is further refined by narrowing the choice to a specific series, such as ``Coleman Sundome,'' that matches both the intent and quality expectation.  

Motivated by these observations, we introduce \textbf{Mind}-inspired \textbf{Rec}ommender (MindRec) to overcome the limitations of auto-regressive mechanisms in recommendation tasks. Our approach leverages Large Language Diffusion with masking (LLaDA) \cite{nie2025llada}, a novel architecture that emulates the generative dynamics of diffusion models. 
Specifically, we first generate key tokens that capture the essential aspects of the target item, then progressively refine these tokens into a complete item representation. To further establish a structured preference refinement process that emulates human decision-making, we organize items within a hierarchical category tree, which allows the model to first generate coarse-grained categories and then gradually narrow down to finer-grained subcategories before generating the specific item. For semantic richness and consistent representation, we represent items using fixed-length semantic IDs from prior work \cite{wang2024learnable,zheng2024adapting}. Finally, to address the local optimum problem inherent in greedy decoding, we design a novel beam search algorithm, Diffusion Beam Search, specifically tailored for our mind-inspired generation paradigm.

MindRec is a new generative recommendation paradigm that breaks the limitations of conventional auto-regressive generation by emulating human reasoning processes. It enables flexible, human-like recommendation generation through structured preference learning and mind-inspired decoding, significantly improving both recommendation accuracy and explainability.
The main contributions of this paper are as follows:
\begin{itemize}[leftmargin=*]
\item We identify the key limitation in existing LLM-based generative recommendation systems: the auto-regressive greedy decoding strategy limits the accuracy of recommendations.
\item We propose MindRec, a novel generative recommendation paradigm that emulates human reasoning by first generating keywords and then refining them into complete recommendations.
\item We introduce a hierarchical category structure to establish a coarse-to-fine preference refinement process, and design a tailored beam search algorithm. %
\item Experimental results demonstrate that MindRec achieves an average improvement of 9.5\% in top-1 recommendation performance over state-of-the-art methods, confirming its superior capability in generating accurate and explainable recommendations.
\end{itemize}

\section{Preliminary}
In this section, we first outline the auto-regressive paradigm in LLM-based recommendation systems, then present codebook-based item representations, and finally introduce the diffusion language model.
\subsection{Auto-regressive LLM-based Recommender}
Next-item recommendation is a fundamental task in recommendation systems, where the goal is to predict the next item a user is likely to interact with based on their historical behavior sequence. With the advent of large language models, this task has been reformulated as a conditional text generation problem, leveraging the generative power of LLMs to produce item recommendations directly \cite{bao2023bi,bao2024decoding}.
In such approaches, each item $i \in \mathcal{I}$ can be represented by token sequence $\mathbf{y}^{(i)} = [y_1^{(i)}, y_2^{(i)}, \dots, y_N^{(i)}]$, where each token $y_n^{(i)}$ belongs to the LLM's vocabulary. Given a prompt $\mathbf{x}$ that describes the task and lists the user's historical interactions, the LLM is trained to generate the token sequence of the target item $\mathbf{y}$ in an auto-regressive manner \cite{gao2025flower}.

Formally, the model policy $\pi_\theta$ is trained to maximize the likelihood of the target token sequence conditioned on the prompt and previously generated tokens using standard cross-entropy loss:
\begin{equation}
\mathcal{L} = -\frac{1}{N} \sum_{n=1}^{N} \log \pi_\theta(y_n \mid \mathbf{x}, y_{1:n-1}),
\end{equation}
where $\pi_\theta(y_n \mid \mathbf{x}, y_{1:n-1})$ denotes the probability of generating token $y_n$ given the context.
During inference, the model generates recommendations token-by-token from left to right:
\begin{equation}
y_n \sim \pi_\theta(\cdot \mid \mathbf{x}, y_{1:n-1}),
\end{equation}
continuing until the end-of-sequence token is produced or a maximum length is reached.

While the auto-regressive paradigm has shown effectiveness in many tasks, the greedy decoding strategy and unidirectional logical chain often result in suboptimal recommendations \cite{lin2025order,10.5555/3037062.3037074}, motivating the need for alternative generation paradigms.

\subsection{Code Identifiers for Item Representation}

In LLM-based generative recommendation, a fundamental challenge is how to effectively represent items in a form that can be processed by language models. Early approaches often rely on textual identifiers, which directly use item titles, attributes, or descriptions to represent items \cite{Harte_2023,bao2023bi}. While these identifiers are rich in semantic information, they lack hierarchical semantic structure, ignore collaborative signals, and introduce noisy tokens. A common alternative involves the use of dense embedding vectors derived from collaborative filtering or representation learning \cite{yang2023dreamrec,li2023e4srec,Ren_2024}. While these embeddings effectively capture collaborative signals and latent item relationships, they are continuous and high-dimensional, incompatible with the discrete token space used in language models. To address these challenges, codebook-based identifiers have been introduced \cite{zheng2024adapting,rajput2023tiger,wang2024learnable}. These methods utilize vector quantization techniques to map items into discrete code sequences, effectively capturing both semantic and collaborative information in a structured form.
In our work, we adopt the code identifiers publicly released by the LETTER \cite{wang2024learnable} and LC-Rec \cite{zheng2024adapting} projects.

\begin{figure*}[!t]
  \centering
  \includegraphics[width=\textwidth]{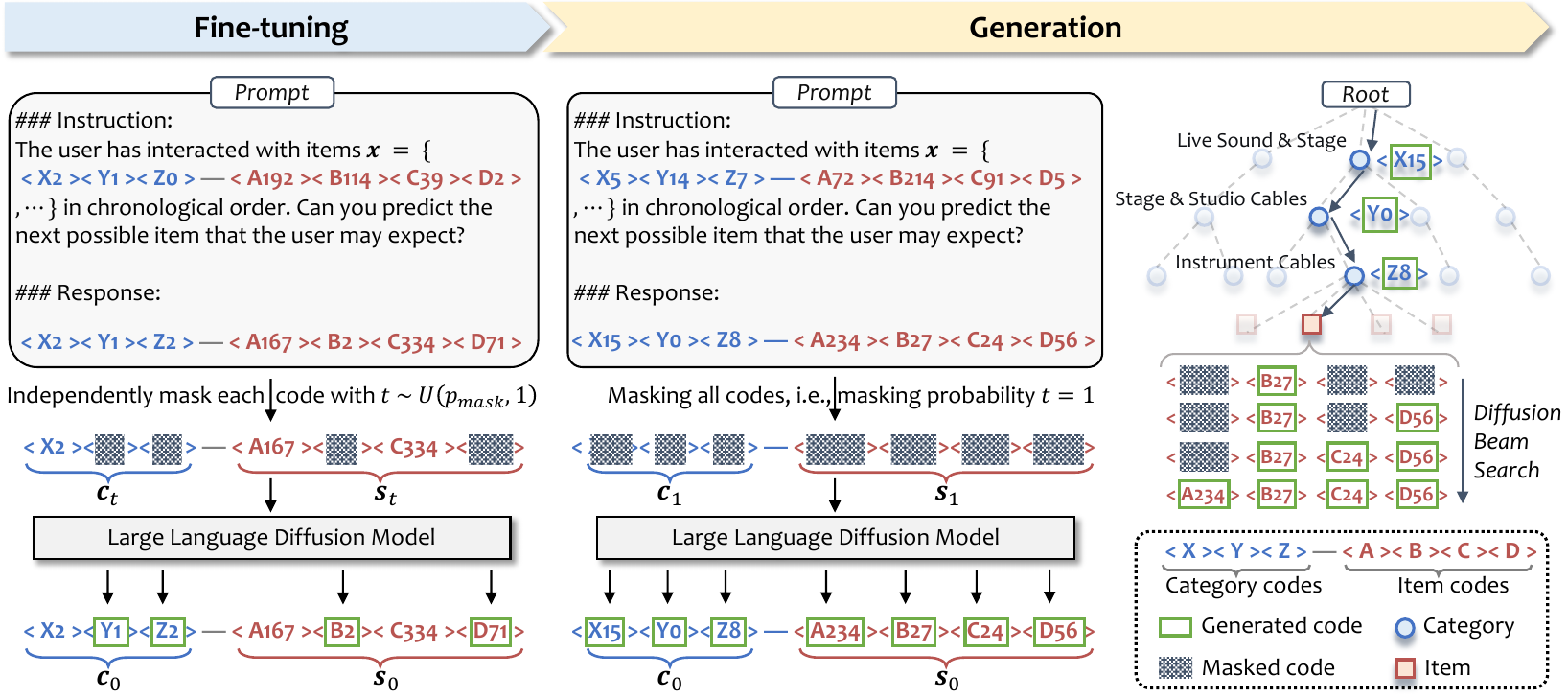}
  \caption{An overview of the MindRec framework.}
  \label{fig:overview}
\end{figure*}

\subsection{Large Language Diffusion Models}
To overcome the inherent limitations of auto-regressive LLMs, we adopt Large Language Diffusion with masking (LLaDA) \cite{nie2025llada}, a masked diffusion model that employs a masking-and-decoding mechanism to simulate the noising and denoising steps. Instead of generating tokens sequentially from left to right, LLaDA learns to reconstruct partially masked sequences.

During pretraining, LLaDA learns to reconstruct the original text from masked versions. For a sequence $x_0$ of length $N$, each token is independently masked with probability $t \sim U[0,1] $, analogous to the noise level in diffusion models, resulting in a masked sequence $x_t$. The model is trained to minimize the cross-entropy loss only over the masked tokens:
\begin{equation}
\mathcal{L}(\theta) = -\mathbb{E}_{t,x_0,x_t} \left[ \frac{1}{t} \sum_{i=1}^{N} \mathbf{1}[x_t^i = \text{M}] \log \pi_\theta(x_0^i | x_t) \right],
\end{equation}
where $x_0^i$ denotes the $i$-th token of $x_0$, $\mathbf{1}[x_t^i = \text{M}]$ is an indicator function that equals 1 if the $i$-th token of $x_t$ is the mask token M, and 0 otherwise.

For instruction following, LLaDA is fine-tuned on prompt-response pairs $(p_0, r_0)$. The prompt $p_0$ remains unmasked, while tokens in the response $r_0$ are masked independently with probability $t$ to produce a masked version $r_t$. The loss is computed as:
\begin{equation}
\mathcal{L}_{\text{SFT}}(\theta) = -\mathbb{E}_{t,p_0,r_0,r_t} \left[ \frac{1}{t} \sum_{i=1}^{N'} \mathbf{1}[r_t^i = \text{M}] \log \pi_\theta(r_0^i | p_0, r_t) \right],
\end{equation}
where $N'$ is the length of the response, $r_0^i$ and $r_t^i$ denote the $i$-th token of $r_0$ and $r_t$, respectively.

During inference, LLaDA generates text by simulating a reverse diffusion process. Starting from a fully masked response sequence, the model iteratively predicts and fills masked tokens. At each step, the model predicts all masked tokens simultaneously and then remasks a fraction of them based on a strategy such as low-confidence remasking. The process continues until all tokens are uncovered.

\section{Method: MindRec}
In this section, we introduce the overall framework of MindRec and detail how it leverages structured preference learning and mind-inspired generation to emulate human reasoning for improved recommendations. An overview is illustrated in \myfig{overview}.

\subsection{Problem Formulation}

Let $\mathcal{I}$ denote the set of all items in the dataset. Each item $i \in \mathcal{I}$ is defined as a tuple of its category and semantic ID, i.e., $i$ = (\textit{\textbf{c}}, \textit{\textbf{s}}) = $([c_1, c_2, \dots, c_L], [s_1, s_2, \dots, s_M])$, where:
\begin{itemize}[leftmargin=*]
\item \textit{\textbf{c}} = $[c_1, c_2, \dots, c_L]$ is a hierarchical category path of length $L$, where each category $c_l$ (for $l > 1$) is a sub-category of the immediately preceding category $c_{l-1}$.
\item \textit{\textbf{s}} = $[s_1, s_2, \dots, s_M]$ is a semantic ID of length $M$.
\end{itemize}
Each data instance consists of a user's interaction history and the next item the user interacts with.
Formally, an instance is a pair (\textit{\textbf{x}}, $y$), where:
\begin{itemize}[leftmargin=*]
\item \textit{\textbf{x}} = $[i_1, i_2, \dots, i_{H}]$ is the user's historical interaction sequence of length $H$, with each $i_h \in \mathcal{I}$.
\item $y \in \mathcal{I}$ is the target item, representing the next item the user interacts with.
\end{itemize}
The model's task is to predict the target item $y$ based on the historical sequence \textit{\textbf{x}}, which is formulated as a conditional generation problem: given the context of historical interactions, the model is prompted to generate the complete structured representation of the next item, including both its hierarchical category and semantic ID.

\subsection{Structured Preference Learning}
\subsubsection{Aligning LLaDA for Recommendation}
Formally, each training instance consists of the user's history \textit{\textbf{x}} and the target item token sequence $y_0$ of fixed length $N$. We obtain a masked sequence ${y}_{t}$ by independently masking each token in $y$ with a probability $t$, where $t$ is sampled from a uniform distribution $U[p_{\text{mask}}, 1]$. This guarantees a minimum masking ratio of $p_{\text{mask}}$, ensuring a non-trivial reconstruction task to promote robust contextual reasoning.
The model is trained to reconstruct the original tokens at all masked positions using the provided context. The training objective is the cross-entropy loss calculated exclusively over the masked tokens:
\begin{equation}
\mathcal{L}(\theta) = -\mathbb{E}_{t,\textit{\textbf{x}},y_0,y_t} \left[ \frac{1}{t} \sum_{i=1}^{N} \mathbf{1}[y_t^i = \text{M}] \log \pi_\theta(y_0^i | \textit{\textbf{x}}, y_t) \right],
\label{eq:aligning} 
\end{equation}
where $y_0^i$ denotes the $i$-th token of $y_0$, $\mathbf{1}[y_t^i = \text{M}]$ is an indicator function that equals 1 if the $i$-th token of $y_t$ is the mask token M, and 0 otherwise. This objective enhances the model's ability to infer user preferences and item relevance by leveraging bidirectional context, which is crucial for accurate recommendations.
\subsubsection{Hierarchical Category Structure}
We utilize three Amazon review datasets, including Arts, Video Games, and Instruments. Each item in these datasets is associated with a hierarchical category list (e.g., ``Instruments, Audio Equipment, Microphone Cables''). We construct a unified category tree by parsing these lists and removing duplicate or redundant entries, ensuring that each node is unique and every item's category list corresponds to a distinct path in the tree.
To convert category paths into machine-processable symbols, we introduce a level-aware coding scheme, in which the code sequence for a category at level $l$ is defined as:
\begin{equation}
\text{code}^{(l)} = \text{code}^{(l-1)} \oplus \langle {\text{prefix}}_l : \text{id} \rangle,
\end{equation}
where $\text{code}^{(l-1)}$ denotes the code sequence of the parent category at level $l-1$, the operator $\oplus$ denotes sequence concatenation, $\text{prefix}_l$ is a fixed symbol representing the level $l$, and $\text{id}$ is a unique identifier within level $l$ that distinguishes the category among its siblings.
For example, in \myfig{overview}, the code sequence ``$\langle X15\rangle\langle Y0\rangle$'' for ``Stage \& Studio Cables'' is formed by concatenating the parent category's code ``$\langle X15\rangle$'' with the $\text{prefix}_2$ ``$Y$'' and the unique identifier ``0'', which distinguishes ``Stage \& Studio Cables'' among its siblings under ``Live Sound \& Stage''.
Each item’s category path is thus mapped to the code sequence of its last category.
Since the length of code sequences varies among items, we analyze their distribution and set a uniform length $L$ to preserve essential information while minimizing noise introduced by artificial padding or truncation. In experiments, we set $L$ to 3 for the Instruments and Arts datasets and to 2 for the Games dataset. Shorter sequences are padded with non-semantic placeholders, while longer ones are truncated to the first $L$ codes.

\subsubsection{Structure Learning Objective}
Based on this processing, we define the hierarchical category sequence for each item as \textit{\textbf{c}} = $\text{code}^{(L)}$ = $[c_1, c_2, \dots, c_L]$, where $c_l$ denotes the code at level $l$. For example, as shown in \myfig{overview}, the category sequence for an item can be represented as ``$\langle X2\rangle\langle Y1\rangle\langle Z0\rangle$'', where each element corresponds to a code at a specific level of the category tree. To capture fine-grained item semantics beyond category information, we incorporate semantic ID sequences of length 4 from prior works \cite{wang2024learnable,zheng2024adapting}, defined as \textit{\textbf{s}} = $[s_1, s_2, s_3, s_4]$ (e.g.,  ``$\langle A192\rangle\langle B114\rangle\langle C39\rangle\langle D2\rangle$'' in \myfig{overview}). The complete item representation is constructed as the concatenation of the category sequence and semantic ID sequence: $[\textit{\textbf{c}},\textit{\textbf{s}}]$, resulting in a unified representation such as ``$\langle X2\rangle\langle Y1\rangle\langle Z0\rangle-\langle A192\rangle\langle B114\rangle\langle C39\rangle\langle D2\rangle$''.

Thus, training objective \eqref{eq:aligning} can be reformulated as follows:
\begin{equation}
\begin{aligned}
\mathcal{L}(\theta) = & -\mathbb{E}\left[ \frac{1}{t} \sum_{i=1}^{L} \mathbf{1}[\textit{\textbf{c}}^i_t = \text{M}] \log \pi_\theta(\textit{\textbf{c}}^i_0 | \textit{\textbf{x}}, \textit{\textbf{c}}_t,\textit{\textbf{s}}_t) \right] \\
                      & -\mathbb{E}\left[ \frac{1}{t} \sum_{i=1}^{4} \mathbf{1}[\textit{\textbf{s}}^i_t = \text{M}] \log \pi_\theta(\textit{\textbf{s}}^i_0 | \textit{\textbf{x}}, \textit{\textbf{c}}_t,\textit{\textbf{s}}_t) \right]
\end{aligned}
\end{equation}
where $\textit{\textbf{c}}_t$ and $\textit{\textbf{s}}_t$ are obtained by randomly masking tokens in original sequence $\textit{\textbf{c}}_0$ and $\textit{\textbf{s}}_0$ (that is, $\textit{\textbf{c}}$ and $\textit{\textbf{s}}$), respectively.
As illustrated in \myfig{overview}, for the sequence ``$\langle X2\rangle\langle Y1\rangle\langle Z2\rangle-\langle A167\rangle\langle B2\rangle\langle C334\rangle\langle D71\rangle$'', each code is independently masked with $t$ sampled from a uniform distribution $U[p_{\text{mask}}, 1]$, resulting in ``$\langle X2\rangle\langle MASK\rangle\langle MASK\rangle-\langle A167\rangle\langle MASK\rangle\langle C334\rangle\langle MASK\rangle$''.

\subsection{Mind-inspired Coarse-to-fine Generation}
\subsubsection{Generation Process}
Initially, the entire token sequence for the target item, including the hierarchical category codes and the specific semantic ID, is fully masked. The model first generates category tokens auto-regressively from the coarsest to the finest level, simulating a top-down reasoning process that progressively narrows the categorical scope before specifying the exact item. This establishes a high-level contextual framework for the subsequent token generation.
Once the hierarchical category path is established, the model generates the semantic ID of the item using the proposed Diffusion Beam Search in a non-autoregressive manner. Rather than following a fixed left-to-right order, this approach dynamically prioritizes tokens based on the model's confidence, allowing early identification and fixation of pivotal elements. These elements then guide the contextualization of the remaining tokens, enabling flexible and human-like item completion.

\medskip \noindent
\textbf{Remark:}
This two-phase generation captures two complementary aspects of human reasoning. The coarse-to-fine category generation resembles a structured, top-down decision-making process \cite{kahneman2011thinking, rosch2024principles}. In parallel, the non-autoregressive item code completion reflects the intuitive process of developing a full idea from an initial spark of inspiration \cite{duncker1945problem, Guilford1950Creativity}. Together, these mind-inspired strategies enhance the accuracy and explainability of the recommendations.
\subsubsection{Diffusion Beam Search}
Our proposed Diffusion Beam Search consists of two main stages at each decoding step: Expansion and Pruning. %
At each step, for each beam (i.e., sequence with masked tokens) in the current beam set $\mathcal{B}$, we compute the probability distribution and extract the top-$b$ candidate tokens for each masked position. These tokens and their corresponding probabilities are aggregated, from which the global top-$b$ tokens are then selected. Each beam is then expanded by substituting the corresponding masked position with its global top-$b$ tokens, generating a new candidate set $\mathcal{C}$.
After expansion, each candidate sequence in $\mathcal{C}$ is evaluated using a diversity-penalized scoring function. The score of a sequence is adjusted based on its similarity to sequences already chosen for the new beam set. Specifically, we define the maximum token overlap rate $\rho$ as the highest proportion of tokens shared with any selected sequence, normalized by the number of decoded tokens at the current step. The adjusted score $f'$ is a penalized version of the original score $f$:
\begin{equation}
f' = f \times \left(1 - \log(1 + \rho)\right)
\end{equation}
This penalty discourages the inclusion of highly similar sequences. We then update $\mathcal{B}_{\text{new}}$ by iteratively adding candidates with the highest adjusted scores while maintaining a maximum of $b$ sequences..
The complete procedure is summarized in Algorithm~\ref{alg:diffusion_beam_search}.

\section{Experiments}
\label{sec:experiment}
In this section, we perform a series of experiments to address the following research questions:

\begin{itemize}[leftmargin=*]
    \item \textbf{RQ1}: How do generative methods perform in the next-item recommendation task?
    \item \textbf{RQ2}: How does structured preference learning with hierarchical categories influence recommendations?
    \item \textbf{RQ3}: What are the effects of the key factors in MindRec?
    \item \textbf{RQ4}: How does the beam search algorithm impact recommendation performance?
\end{itemize}

\subsection{Experimental Setup} 

\begin{table}[!t] 
\centering %
\caption{Dataset statistics before and after processing.}
\begin{tabular}{ccccc}
\toprule
\multicolumn{2}{c}{Dataset}                                                                                                       & \#User & \#Item & \#Interaction \\ \midrule
\multicolumn{1}{c}{\multirow{2}{*}{\begin{tabular}[c]{@{}c@{}}Instruments\end{tabular}}} & Before & 24771 & 9921 & 206153        \\
\multicolumn{1}{c}{}                                                                                         & After  & 5000   & 9223   & 56001         \\ \midrule
\multicolumn{1}{c}{\multirow{2}{*}{\begin{tabular}[c]{@{}c@{}}Arts\end{tabular}}}   & Before & 45140 & 20955  & 390832        \\
\multicolumn{1}{c}{}                                                                                         & After  & 5000   & 17054   & 63401         \\ \midrule
\multicolumn{1}{c}{\multirow{2}{*}{\begin{tabular}[c]{@{}c@{}}Games\end{tabular}}} & Before & 204439 & 16858  & 452989        \\
\multicolumn{1}{c}{}                                                                                         & After  & 5000   & 14671   & 77096         \\ \bottomrule
\end{tabular}
\label{tab:Statistic Information} %
\end{table}

\subsubsection{Dataset}
Following the previous research \cite{wang2024learnable,zheng2024adapting}, we conduct experiments on three real-world datasets
from Amazon review data\footnote{https://cseweb.ucsd.edu/\textasciitilde jmcauley/datasets.html$\#$amazon$\_$reviews}
, including Arts, Video Games, and
Instruments. These datasets contain user review data from May 1996 to October 2018 and provide detailed item information, including title, brand, category, and description.
For efficiency, based on publicly available datasets from prior work \cite{wang2024learnable,zheng2024adapting}, we select 5,000 users and limit the maximum historical sequence length to 10. We adopt the leave-one-out strategy, where the last interaction of each user is held out for testing, the second-to-last for validation, and all preceding interactions for training. Detailed statistical information is provided in \mytab{Statistic Information}.

\subsubsection{Evaluation Protocal}
The performance evaluation of top-$k$ recommendation encompasses both accuracy and diversity.

For \textbf{Accuracy}, following prior work \cite{bao2023bi,bao2024decoding,jiang2024item}, we adopt two widely used metrics: Hit Ratio (HR@K) and Normalized Discounted Cumulative Gain (NDCG@K).  
To assess \textbf{Diversity}, we use two metrics: (1) Entropy of Codes (Entropy), which calculates the entropy of all codes in the recommended items, and (2) Type-Token Ratio (TTR), defined as the ratio of unique codes to the total number of codes in the recommendations.

In the comparison results, we report HR@1 (same as NDCG@1), HR@3, and NDCG@3 as evaluation metrics. For the ablation studies, we also examine the diversity of recommendations, evaluating the results with both Entropy and TTR.

\subsubsection{Baseline}
We select one traditional sequential recommendation model and several LLM-based generative recommendation methods as baselines:
\begin{itemize}[leftmargin=8pt]
    \item \textbf{SASRec} \cite{kang2018self} is a widely used sequential recommendation baseline employing a self-attention mechanism.
    \item \textbf{Tiger} \cite{rajput2023tiger} is one of the earliest and most classic methods that introduce codebook-based identifiers via RQ-VAE to quantize semantic information into a code sequence for LLM-based generative recommendation.
    \item \textbf{IDGenRec} \cite{tan2024idgenrec} is a LLM-based framework that learns to represent items as concise, semantically-rich textual IDs using natural language tokens.
    \item \textbf{LETTER} \cite{wang2024learnable} is a learnable tokenizer that integrates hierarchical semantics, collaborative signals, and code assignment diversity into item identifiers for generative recommendation. We report the performances of LETTER instantiated on LC-Rec \cite{zheng2024adapting} with various base models.
\end{itemize}

\subsubsection{Implementation Details}
For SASRec, we optimize using binary cross-entropy loss and the Adam optimizer, with a learning rate in [1e-3, 2e-3], a batch size of 256, and maximum training epochs of 200. 
In our experiments, the large language models employed include Llama-3-8B-Instruct, Qwen2.5-7B-Instruct, and LLaDA-8B-Instruct.
To ensure fairness across LLM-based methods, all approaches adopt the same code identifiers for item representation.
For Tiger, we set the learning rate to [5e-4, 1e-3], with a batch size of 1024 and a maximum of 200 training epochs.
For IDGenRec, the model is trained over 3 rounds. In each round, the ID generator is trained for 1 epoch and the recommender for 10 epochs, using learning rates of 1e-3 and 1e-8, respectively.
For the auto-regressive LLM-based implements of LETTER, the learning rate is chosen from [5e-5, 1e-4, 2e-4], while for the LLaDA-based LETTER and our proposed method, we use learning rates in [5e-5, 7e-5, 1e-4] and a lower-bound mask ratio $p_{\text{mask}}$ in [0.4, 0.5, 0.6, 0.7].
Both LETTER and MindRec are trained for 3 epochs.

\begin{table*}[]
\caption{Performance of all methods on three real-world datasets. The best results are bolded.}
\begin{tabular}{c|ccc|ccc|ccc}
\hline
                                & \multicolumn{3}{c|}{Instruments}                    & \multicolumn{3}{c|}{Arts}                           & \multicolumn{3}{c}{Games}                           \\ \hline
                                & \textbf{HR@1}$\uparrow$   & \textbf{NDCG@3}$\uparrow$ & \textbf{HR@3}$\uparrow$   & \textbf{HR@1}$\uparrow$   & \textbf{NDCG@3}$\uparrow$ & \textbf{HR@3}$\uparrow$   & \textbf{HR@1}$\uparrow$   & \textbf{NDCG@3}$\uparrow$ & \textbf{HR@3}$\uparrow$   \\ \hline
SASRec                          & 0.0210          & 0.0335          & 0.0428          & 0.0114          & 0.0182          & 0.0232          & 0.0050          & 0.0089          & 0.0118          \\
Tiger                           & 0.0386          & 0.0431          & 0.0468          & 0.0144          & 0.0190          & 0.0228          & 0.0068          & 0.0121          & 0.0160          \\
IDGenRec                        & 0.0432          & 0.0493          & \textbf{0.0540} & 0.0252          & 0.0304          & 0.0340          & 0.0074          & 0.0126          & 0.0162          \\
LETTER-Llama3                          & 0.0372          & 0.0425          & 0.0464          & 0.0220          & 0.0263          & 0.0294          & 0.0060          & 0.0089          & 0.0108          \\
LETTER-Qwen2.5 & 0.0408          & 0.0483          & 0.0538          & 0.0230          & 0.0274          & 0.0304          & 0.0082          & 0.0109          & 0.0130          \\
LETTER-LLaDA   & 0.0452          & 0.0498          & 0.0530          & 0.0230          & 0.0269          & 0.0298          & 0.0060          & 0.0092          & 0.0116          \\  \hline
\rowcolor[HTML]{EFEFEF}
MindRec  & \textbf{0.0464} & \textbf{0.0506} & 0.0536          & \textbf{0.0280} & \textbf{0.0315} & \textbf{0.0342} & \textbf{0.0094} & \textbf{0.0135} & \textbf{0.0166} \\ \hline
\end{tabular}
\label{tab:main result}
\end{table*}

\subsection{Next-item Recommendation Results (RQ1)}  
\label{sec:RQ1}
We evaluate the recommendation performance of MindRec and baseline methods on the next-item generative recommendation task using three open-world datasets. The overall experimental results are summarized in \mytab{main result}, and the key observations are:
\begin{itemize}[leftmargin=*]
    \item MindRec outperforms all baselines across three datasets, except for a slight shortfall in HR@3 on the Instruments dataset compared to IDGenRec. These results indicate the superiority of our framework as a novel paradigm for generative recommendation.
    \item MindRec consistently surpassed other LLM-based generative methods, indicating that the conventional auto-regressive generation inherently limits recommendation performance. In contrast, our proposed mind-inspired generation effectively overcomes these limitations, offering a more flexible, accurate, and human-like paradigm.
    \item LLaDA-based implementation of LETTER outperforms LETTERs built on other large language models across all evaluation metrics under identical experimental settings. This finding further validates the advantages of the mind-inspired generation, highlighting its general applicability and effectiveness.
    \item When comparing LLaDA-based implementations, MindRec's advantage over LETTER-LLaDA underscores the importance of structured preference learning and Diffusion Beam Search. The coarse-to-fine decoding emulates human decision-making, while the novel beam search algorithm mitigates the issue of suboptimal sequences and improves diversity, yielding more accurate and explainable recommendations.
\end{itemize}

\begin{table*}[]
\caption{Ablation study on the use of hierarchical category structure during training and inference.}
\resizebox{\textwidth}{!}{%
\begin{tabular}{c|ccccccccc}
\hline
                         & \textbf{HR@1}$\uparrow$   & \textbf{NDCG@3$\uparrow$} & \textbf{HR@3$\uparrow$}   & \textbf{NDCG@5}$\uparrow$ & \textbf{HR@5}$\uparrow$   & \textbf{NDCG@10}$\uparrow$ & \textbf{HR@10}$\uparrow$  & \textbf{Entropy}$\uparrow$ & \textbf{TTR}$\uparrow$    \\ \hline
Training w/o category      & 0.0230          & 0.0269          & 0.0298          & 0.0280          & 0.0324          & 0.0287           & 0.0348          & 8.950            & 0.0093          \\ \hline
Inference w/o category  & 0.0238          & 0.0287          & 0.0322          & 0.0297          & 0.0346          & 0.0299           & 0.0352          & 8.610            & 0.0048          \\
Inference with category  & 0.0280          & 0.0315          & 0.0342          & 0.0325          & 0.0364          & 0.0328           & 0.0372          & 8.738            & 0.0092          \\
Inference given category & \textbf{0.0546} & \textbf{0.0673} & \textbf{0.0760} & \textbf{0.0708} & \textbf{0.0844} & \textbf{0.0719}  & \textbf{0.0880} & \textbf{9.191}   & \textbf{0.0096} \\ \hline
\end{tabular}
}
\label{tab:category}
\end{table*}

\subsection{Role of Structured Preference (RQ2)}  
\label{sec:RQ3}  
To evaluate the role of structured preference learning, particularly the effect of incorporating hierarchical categories, we conduct a series of ablation studies on the Arts dataset under four training and inference settings:
(1) Training without category: Hierarchical categories are entirely omitted during both training and inference.
(2) Inference without category: Hierarchical categories are included in the user’s interaction history, but only the item's semantic ID is generated.
(3) Inference with category: This setting aligns with our main experiments. The model is trained with hierarchical categories and generates both the category path and semantic ID, simulating a coarse-to-fine reasoning process.
(4) Inference given category: The model is trained with hierarchical categories, and the ground-truth category path is provided during inference; only the semantic ID is generated.

As shown in \mytab{category}, using hierarchical categories during training consistently improves accuracy, indicating that structured category sequences help the model better capture item relationships and user preferences. When categories are incorporated during inference through coarse-to-fine generation, both accuracy and diversity improve over ``Inference without categories'' setting, suggesting that explicitly generating category paths helps narrow down the candidate space in a structured manner, reducing ambiguity and reinforcing decision confidence.
Notably, the ``Inference given categories'' setting yields the strongest performance, underscoring the critical role of accurate categorical context in semantic ID generation. These findings confirm that hierarchical categories serve as a foundational component of structured preference learning, enabling the generation of more precise and human-like recommendations.

\subsection{Analysis of Key Factors in MindRec (RQ3)}  
\label{sec:RQ4}  
To comprehensively assess the effectiveness of the proposed MindRec framework and quantify the contribution of each core component, we perform a series of ablation studies. The following sections examine the impact of four key factors: model size, decoding strategy, the lower-bound masking ratio $p_{\text{mask}}$ used during training, and the beam size employed during inference. Each factor plays a critical and distinct role in model performance: (1) model scale governs representation capacity; (2) the decoding strategy influences the flexibility and optimality of generation; (3) $p_{\text{mask}}$ controls the model's generalization and contextual reasoning ability; and (4) beam size balances search breadth against computational efficiency.
These experiments are designed to isolate the effects of individual elements and provide deeper insights into the model's behavior and robustness.

\begin{table*}[]
\caption{Performance of MindRec with different model sizes on the Arts dataset.}
\resizebox{\textwidth}{!}{%
\begin{tabular}{c|ccccccccc}
\hline
                                        & \textbf{HR@1}$\uparrow$ & \textbf{NDCG@3}$\uparrow$ & \textbf{HR@3}$\uparrow$ & \multicolumn{1}{c}{\textbf{NDCG@5}$\uparrow$} & \multicolumn{1}{c}{\textbf{HR@5}$\uparrow$} & \multicolumn{1}{c}{\textbf{NDCG@10}$\uparrow$} & \multicolumn{1}{c}{\textbf{HR@10}$\uparrow$} & \textbf{Entropy}$\uparrow$ & \textbf{TTR}$\uparrow$ \\ \hline
LLaDA-1.5B     &  0.0274        & 0.0301          & 0.0320        & 0.0310                              & 0.0340                            & 0.0315                               & 0.0358                             & \textbf{8.765}            & 0.0092       \\
LLaDA-8B-Instruct & \textbf{0.0280}        & \textbf{0.0315}          & \textbf{0.0342}        & \textbf{0.0325}                              & \textbf{0.0364}                            & \textbf{0.0328}                               & \textbf{0.0372}                             & 8.738            & \textbf{0.0092}       \\ \hline
\end{tabular}
}
\label{tab:model parameter}
\end{table*}

\subsubsection{Impact of Model Parameter}  
\label{sec:RQ41} 
In our main experiments, we adopt LLaDA-8B-Instruct as the base model. To further investigate the impact of model size on recommendation performance, we implement the MindRec framework using LLaDA-1.5B and evaluate it on the Arts dataset. All experimental configurations remained consistent with the main experiments except for the base model. As summarized in \mytab{model parameter}, the results show that while LLaDA-1.5B achieves diversity performance comparable to LLaDA-8B-Instruct, the latter demonstrates superior accuracy across all top-$k$ metrics. These findings suggest that model sizes play a crucial role in recommendation accuracy, while diversity remains relatively stable across different parameter sizes.

\begin{figure}[!t]
\setlength{\abovecaptionskip}{0.2cm}
  \centering
  \includegraphics[width=\columnwidth]{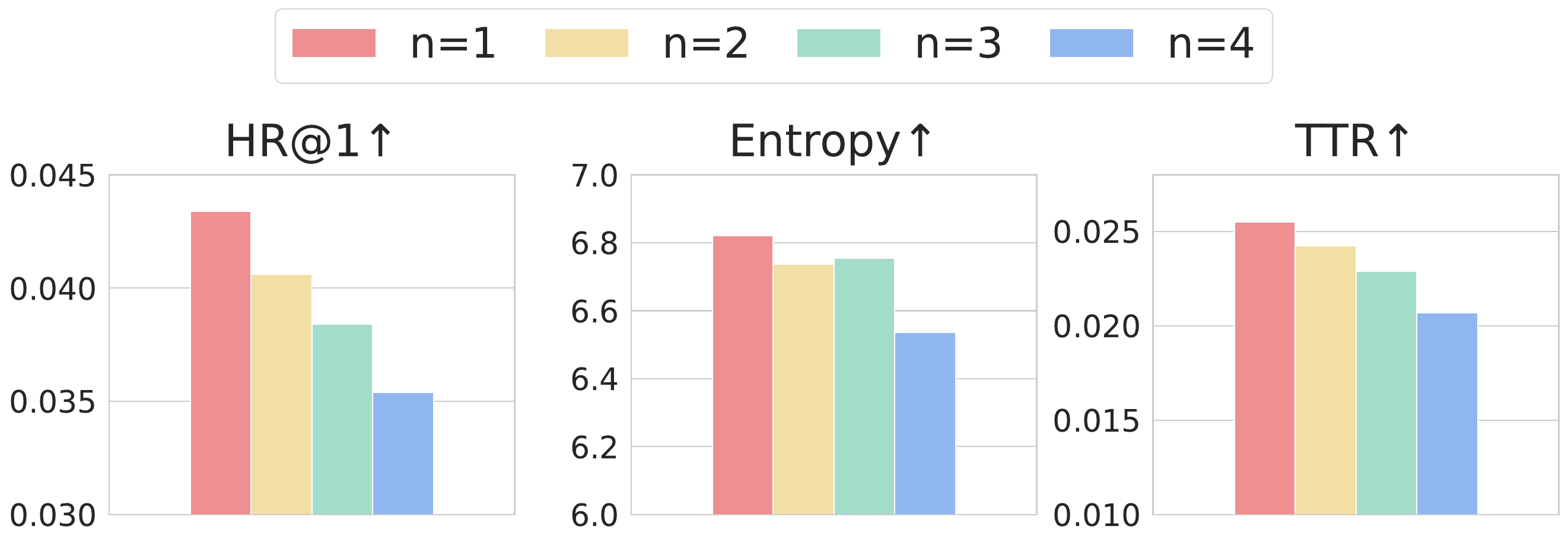} %
  \vspace{-1mm}
  \caption{Recommendation Results of decoding $n$ tokens per step on the Instruments dataset.} %
  \vspace{-1mm}
  \label{fig:token_per_step} %
\end{figure}

\begin{table*}[]
\caption{Impact of decoding flexibility on recommendation performance, where larger $m$ indicates greater flexibility.}
\resizebox{\textwidth}{!}{%
\begin{tabular}{c|ccccccccc}
\hline
                                        & \textbf{HR@1}$\uparrow$ & \textbf{NDCG@3}$\uparrow$ & \textbf{HR@3}$\uparrow$ & \multicolumn{1}{c}{\textbf{NDCG@5}$\uparrow$} & \multicolumn{1}{c}{\textbf{HR@5}$\uparrow$} & \multicolumn{1}{c}{\textbf{NDCG@10}$\uparrow$} & \multicolumn{1}{c}{\textbf{HR@10}$\uparrow$} & \textbf{Entropy}$\uparrow$ & \textbf{TTR}$\uparrow$ \\ \hline
$m$=1                             & 0.0260        & 0.0298          & 0.0326        & 0.0308                              & 0.0352                            & 0.0320                               & \textbf{0.0390}                             & 8.621            & 0.0088       \\
$m$=2     & 0.0270        & 0.0309          & 0.0336        & 0.0320                              & 0.0362                            & 0.0327                               & 0.0386                             & 8.631            & 0.0090       \\
$m$=4 & \textbf{0.0280}        & \textbf{0.0315}          & \textbf{0.0342}        & \textbf{0.0325}                              & \textbf{0.0364}                            & \textbf{0.0328}                               & 0.0372                             & \textbf{8.738}            & \textbf{0.0092}       \\ \hline
\end{tabular}
}
\label{tab:n_token_as_a_block}
\end{table*}

\subsubsection{Effects of Decoding Strategy}  
\label{sec:RQ42}
The experiments described above are conducted under the assumption that only one token is decoded per step, while the decoding order may vary flexibly across the target item's token sequence. In practice, however, it is feasible to decode multiple tokens per step or impose specific constraints on the decoding order.

We first evaluate the effect of decoding $n$ tokens per step on the Instruments dataset. Since each semantic ID has a fixed length of 4, we test $n$ = 1, 2, 3, and 4. When $n$ = 4, the entire item is generated in one step. However, this parallel decoding approach significantly increases the complexity of the generation process. During beam search, identifying the optimal top-$k$ token combinations becomes computationally challenging. Thus, we only report the top-1 recommendation results in this setting. 
As shown in \myfig{token_per_step}, both accuracy and diversity improve as $n$ declines, indicating that a more careful and fine-grained decoding process leads to better recommendation performance.
We also experiment with constrained decoding order on the Arts dataset by grouping tokens into blocks of size $m$, where all tokens in a block must be decoded before advancing to the next. When $m$ = 1, the process resembles auto-regressive generation. The results in \mytab{n_token_as_a_block} show that as $m$ increases, the constraints are progressively relaxed, enhancing the flexibility of the generation process. This leads to improvements in both accuracy and diversity, further underscoring the flexibility and effectiveness of mind-inspired generation.

\subsubsection{Performance Varying $p_{\text{mask}}$} 
\label{sec:RQ43}  

\begin{figure}[!t]
\setlength{\abovecaptionskip}{0.2cm}
  \centering
  \includegraphics[width=\columnwidth]{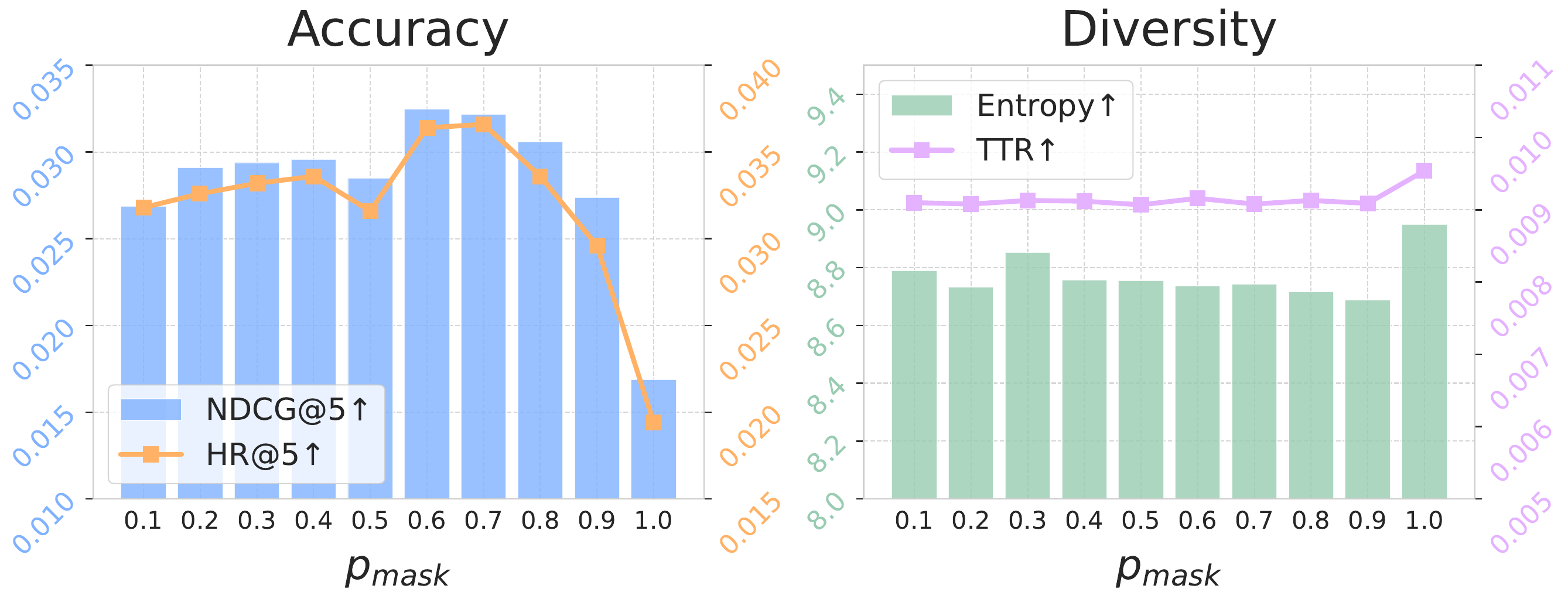} %
  \vspace{-1mm}
  \caption{Performance with varying $p_{\text{mask}}$ on the Arts dataset.} %
  \vspace{-1mm}
  \label{fig:pmask} %
\end{figure}
The masking ratio employed during training critically influences the model's ability to reconstruct the original sequence from a partially masked input, prompting us to investigate its impact. Experimental results on the Arts dataset are reported in \myfig{pmask}. As $p_{\text{mask}}$ increases from left to right, accuracy first rises and then declines, while diversity remains relatively stable.
When $p_{\text{mask}}$ is low, the reconstruction task becomes trivial since only a small portion of tokens are masked. In this case, the model fails to develop robust contextual reasoning capabilities, leading to suboptimal accuracy. Conversely, when $p_{\text{mask}}$ is high, the input sequence is almost entirely masked, leaving insufficient contextual information for meaningful inference. This overwhelms the model and degrades its reconstruction performance, thereby reducing accuracy.
The trade-off is achieved when $p_{\text{mask}}$ is around 0.6, yielding the optimal overall recommendation accuracy.

\begin{figure}[!t]
\setlength{\abovecaptionskip}{0.2cm}
  \centering
  \includegraphics[width=\columnwidth]{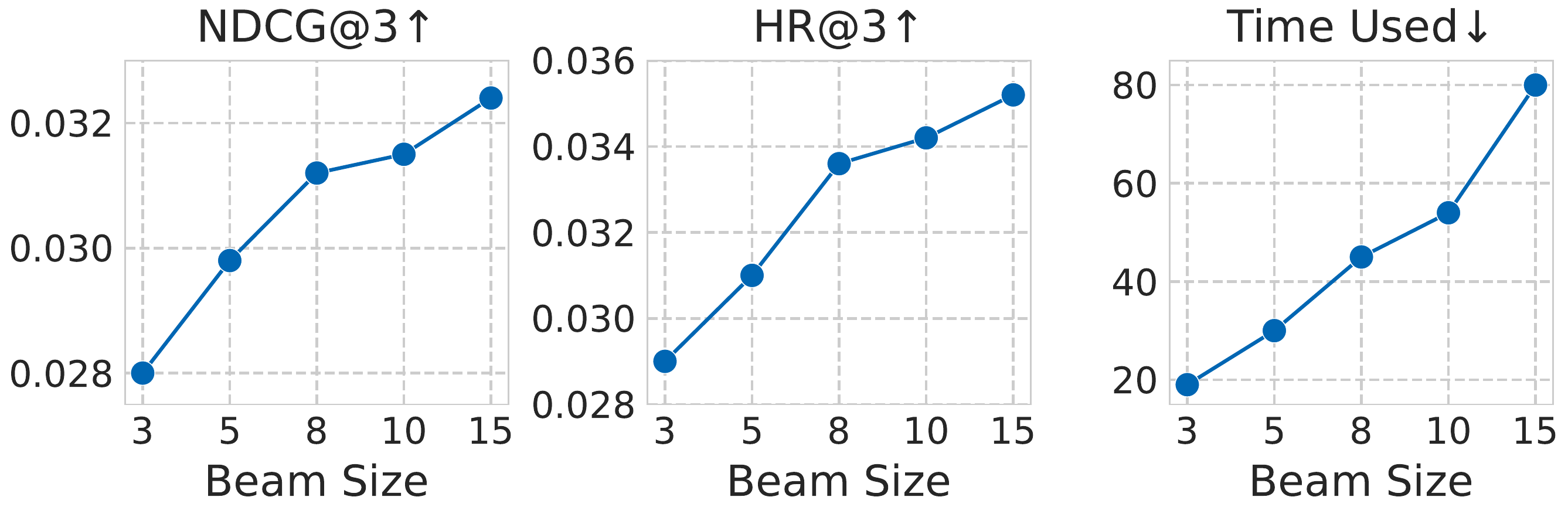} %
  \vspace{-1mm}
  \caption{Performance of MindRec with different beam sizes on the Arts dataset.} %
  \vspace{-1mm}
  \label{fig:beam_size} %
\end{figure}

\begin{table*}[!]
\caption{Performance comparison of different beam search algorithms on the Arts dataset.}
\resizebox{\textwidth}{!}{%
\begin{tabular}{c|ccccccccc}
\hline
                                        & \textbf{HR@1}$\uparrow$ & \textbf{NDCG@3}$\uparrow$ & \textbf{HR@3}$\uparrow$ & \multicolumn{1}{c}{\textbf{NDCG@5}$\uparrow$} & \multicolumn{1}{c}{\textbf{HR@5}$\uparrow$} & \multicolumn{1}{c}{\textbf{NDCG@10}$\uparrow$} & \multicolumn{1}{c}{\textbf{HR@10}$\uparrow$} & \textbf{Entropy}$\uparrow$ & \textbf{TTR}$\uparrow$ \\ \hline
BS                             & 0.0278        & 0.0291          & 0.0300        & 0.0298                              & 0.0318                            & 0.0303                               & 0.0334                             & 8.578            & 0.0086       \\
DBS-Deduplication     & 0.0280        & 0.0313          & 0.0338        & 0.0324                              & 0.0364                            & 0.0327                               & 0.0372                             & 8.737            & 0.0091       \\
DBS-Diversity penalty & \textbf{0.0280}        & \textbf{0.0315}          & \textbf{0.0342}        & \textbf{0.0325}                              & \textbf{0.0364}                            & \textbf{0.0328}                               & \textbf{0.0372}                             & \textbf{8.738}            & \textbf{0.0092}       \\ \hline
\end{tabular}
}
\label{tab:beam search}
\end{table*}

\subsubsection{Impact of Beam Size} 
As shown in \myfig{beam_size}, both NDCG@ and HR@3 improve with increasing beam size, indicating that expanding the beam broadens the search space and enhances the likelihood of generating optimal recommendations. However, this improvement in accuracy comes at the expense of computational efficiency. As the beam size grows, the inference time for recommendation generation increases accordingly, highlighting a trade-off between recommendation quality and computational efficiency.
These results suggest that while larger beam sizes yield more accurate and relevant recommendations, they also demand greater computational resources and longer inference times, which may hinder their applicability in real-time or resource-constrained scenarios.

\subsection{Discussion on Beam Search Algorithm (RQ4)}  
\label{sec:RQ2}  
In this subsection, we evaluate the impact of different beam search algorithms on recommendation performance using the Arts dataset. We compare three variants: (1) Standard beam search, which emulates conventional auto-regressive decoding by selecting globally optimal tokens or beams at each step. (2) Deduplication-based diffusion beam search, which excludes any beam that would generate a sequence identical to one already selected at each decoding step.(3) Diversity penalty-based diffusion beam search, the approach adopted in our main experiments, where a penalty is applied to each beam's score according to its maximum token overlap rate with previously selected beams.

As shown in \mytab{beam search}, the standard beam search yields the worst performance in terms of both accuracy and diversity. Furthermore, increasing the top-$k$ value brings only marginal improvements in accuracy, due to the extremely high repetition rate in generated sequences.
Although a larger $k$ expands the candidate pool, most sequences are identical or nearly identical, leaving the effective search space limited.
The deduplication-based diffusion beam search mitigates this issue by preventing the generation of exactly duplicate sequences, improving both accuracy and diversity. However, it fails to handle highly similar yet non-identical sequences, which still restrict diversity.  In contrast, the diversity-penalized diffusion beam search introduces a soft penalty term based on the maximum token overlap rate with previously selected beams. This discourages the generation of both identical and highly similar sequences, promoting exploration of a broader and more diverse candidate set during decoding, and ultimately achieving the best performance in both accuracy and diversity.

\section{Related Work}
\label{sec:related}
This section reviews the literature relevant to our work, with a focus on LLM-based generative recommendation and diffusion language models. We also highlight the limitations of existing approaches and position our method within the current research landscape.

\subsection{LLMs for Generative Recommendation}
Large Language Models (LLMs) have demonstrated remarkable capabilities in text generation, reasoning, and generalization \cite{zhao2025surveyllm,minaee2025llmsurvey}, which has motivated their adoption in personalized recommendation tasks. In such efforts, a recommendation is formulated as a conditional text generation problem. Specifically, items are represented using textual descriptions (e.g., titles, attributes, or categories) \cite{bao2023bi,zhang2023collm,chen2025dlcrec,liao2024llara} or semantic IDs \cite{rajput2023tiger,zheng2024adapting,wang2024learnable}, and the model is trained to generate the next item based on a user's historical interaction sequence.
Existing LLM-based recommendation methods typically rely on auto-regressive training and generation, where items are predicted token-by-token in a strict left-to-right manner \cite{gao2025flower}. Despite their effectiveness, the greedy decoding strategy and unidirectional logical flow often lead to suboptimal recommendations \cite{lin2025order,10.5555/3037062.3037074}. 
In contrast, MindRec emulates human reasoning by first generating key tokens that reflect user preferences and then refining them into complete items, enabling more flexible and human-like recommendation generation.

\subsection{Diffusion Language Model}
Diffusion models \cite{cao2023surveydiffusion,yang2024diffusionsurvey}, originally developed for continuous data such as images \cite{ho2020DDPM,sohldickstein2015deep,rombach2022high,song2020generative}, have recently been adapted to discrete sequential data like text \cite{austin2023structure, hoogeboom2021argmax}. Early approaches either projected text into continuous embedding spaces \cite{li2022diffusionlm,gong2023diffuseq,li2023RenderDiffusion} or applied parameterized diffusion to discrete distributions \cite{he2022diffusionbert,lou2024discretediffusion,campbell2022continuous}. More recent masked diffusion models (MDMs) replaced the noising process with discrete masking, achieving strong performance in text generation \cite{sahoo2024simple,xu2025energy,austin2023structure}. However, scaling such models to the size of modern large language models remained an open challenge.

Large Language Diffusion with masking (LLaDA) \cite{nie2025llada} represents a significant breakthrough by scaling a diffusion-based language model to 8 billion parameters. Unlike auto-regressive models, LLaDA emulates the generative dynamics of diffusion models, enabling more flexible generation.
In this work, by adopting LLaDA, we propose a new generative recommendation framework that emulates human reasoning processes, significantly improving recommendation performance.

\section{Conclusion}
In this paper, we identify a fundamental limitation in current LLM-based generative recommendation systems: the auto-regressive decoding strategy often leads to suboptimal outcomes, limiting the accuracy of recommendations. To overcome this limitation, we propose MindRec, a novel framework that emulates human reasoning by first generating key tokens that reflect user preferences and then refining them into complete recommendations.
To further emulate structured human decision-making, we organize items into a hierarchical category tree, which guides the model to first generate a coarse-grained category, then progressively refine it through finer-grained subcategories before generating the specific item. To address the local optimum issue inherent in greedy decoding, we introduce Diffusion Beam Search, a novel algorithm tailored for this mind-inspired generation paradigm.
Extensive experiments validate MindRec's effectiveness in generating accurate and explainable recommendations.

MindRec challenges the prevailing auto-regressive paradigm in generative recommendation and proposes a more effective, flexible alternative. By introducing the mind-inspired generation and structured preference learning, MindRec not only achieves strong performance gains but also opens up new avenues for designing more intelligent and human-like recommenders, shifting the design philosophy of recommendation systems from pure mathematical optimization toward human reasoning simulation. Future work may further explore and integrate insights from psychology and biology on human thought processes into the generative framework, enabling more natural, engaging, and trustworthy user interactions.

\bibliographystyle{ACM-Reference-Format}
\balance
\bibliography{Flower}

\clearpage
\appendix

\section{Diffusion Beam Search}
Our proposed Diffusion Beam Search consists of two core procedures: Expansion and Pruning. The Expansion step generates new candidate sequences by decoding the most promising tokens across all masked positions. The Pruning step selects the best sequences while penalizing those that are highly similar to previously selected ones, thereby promoting diversity. The complete procedure is presented as Algorithm~\ref{alg:diffusion_beam_search}.

\begin{algorithm}
\caption{Diffusion Beam Search}
\label{alg:diffusion_beam_search}
\begin{algorithmic}[1]
\Require Initial fully masked sequence $s_0$, beam size $B$, maximum decoding steps $T$
\Ensure Final set of completed sequences $\mathcal{B}$
\State $\mathcal{B} \leftarrow \{(s_0, 1)\}$ 
\For{$step = 1$ to $T$}
    \State $\mathcal{C} \leftarrow \emptyset$ 
    \For{each beam $(b, f_b) \in \mathcal{B}$}
        \For{each $p \in \text{masked positions in } b$}
            \State $\pi_p \leftarrow \text{probability distribution at position } p$
            \State $\mathcal{T}_p \leftarrow \text{Top}_b(\pi_p)$
        \EndFor
        \State $\mathcal{T}_{\text{global}} \leftarrow \text{global top-}b \text{ tokens from all } \mathcal{T}_p$
        \For{each $(tok, pos, prob) \in \mathcal{T}_{\text{global}}$}
            \State $b' \leftarrow \text{fill token } tok \text{ into position } pos \text{ in beam } b$
            \State $f_{b'} \leftarrow f_b \times prob$ 
            \State $\mathcal{C} \leftarrow \mathcal{C} \cup \{(b', f_{b'})\}$
        \EndFor
    \EndFor
    \State Sort $\mathcal{C}$ by $f$ in descending order
    \State $\mathcal{B}_{\text{new}} \leftarrow \emptyset$
    \For{each $(s, f_s) \in \mathcal{C}$}
        \State $\rho \leftarrow \max_{s' \in \mathcal{B}_{\text{new}}} \frac{\text{count\_same\_token}(s, s')}{\text{current\_step}}$
        \State $f'_s \leftarrow f_s \times (1 - \log(1 + \rho))$ 
        \If{$|\mathcal{B}_{\text{new}}| < B$ or $f'_s > \min\{f'_{s'} \mid s' \in \mathcal{B}_{\text{new}}\}$}
            \State Add $(s, f'_s)$ to $\mathcal{B}_{\text{new}}$
            \State Keep top-$B$ sequences in $\mathcal{B}_{\text{new}}$ by $f'$
        \EndIf
    \EndFor
    \State $\mathcal{B} \leftarrow \mathcal{B}_{\text{new}}$
\EndFor
\State \Return $\mathcal{B}$
\end{algorithmic}
\end{algorithm}

\end{document}